\documentclass[twocolumn,showpacs,preprintnumbers,amsmath,amssymb,floatfix]{revtex4-1}
\usepackage{graphicx}% Include figure files
\usepackage{dcolumn}% Align table columns on decimal point
\usepackage{bm}% bold math
\begin{document}
  
%\preprint{APS/123-QED}

\title{Analogue of surface melting in a macroscopic non-equilibrium system}
\author{Christopher May}
\author{Michael Wild}
\author{Ingo Rehberg}
\author{Kai Huang}
\email{kai.huang@uni-bayreuth.de}
\affiliation{Experimentalphysik V, Universit\"at Bayreuth, 95440 Bayreuth, Germany}

\date{\today}% It is always \today, today,

\begin{abstract}

Agitated wet granular matter can be considered as a non-equilibrium model system for phase transitions, where the macroscopic particles replace the molecules and the capillary bridges replace molecular bonds. It is demonstrated experimentally that a two dimensional wet granular crystal driven far from thermal equilibrium melts from its free surface, preceded by an amorphous state. The transition into the surface melting state, as revealed by the bond orientational order parameters, behaves like a first order phase transition, with a threshold being traceable to the rupture energy of a single capillary bridge. The observation of such a transition in the macroscopic non-equilibrium system triggers the question on the universality of surface melting.

\end{abstract}
  
% PACS
   
\maketitle

%%%%%%%%%%%%%%%%%%%%%%%%%%%%%%%%%%%%%%%%%%%%%%%%%%%%%%%
%%%%%%%%  Introduction  %%%%%%%%%%%%%%%%%%%%%%%%%%%%%%%
%%%%%%%%%%%%%%%%%%%%%%%%%%%%%%%%%%%%%%%%%%%%%%%%%%%%%%%

Surface melting has been a topic of interest since Michael Faraday's observations on regelation, the welding of two blocks of ice after contact below $0\,^{\circ}{\rm C}$ \cite{Faraday1850}. After more than a century's investigations, it becomes clear that melting is a continuous process that tends to start from the free surface \cite{Dash99, Tartaglino05, Dash06}. The qualitative idea initiated by Frenkel \cite{Frenkel46} is the reduction of surface energy due to the weaker binding of molecules at the surface compared with that within the bulk. Quantitative experiments pioneered by Frenken and colleagues \cite{Frenken85, Frenken86} have revealed that many solids melt by forming a premelted film, an intermediate state between a solid and a liquid, below the bulk melting temperature \cite{*[][{, and references therein.}] Dash06}. From a microscopic perspective, the kinetics of melting transition has also been explored in detail by means of molecular dynamics simulations \cite{Cahn86, Jin01, Cahn01} and experiments \cite{Zahn00, Gasser2001, Alsayed2005, Gasser10, Wang12} with colloidal suspensions as model systems, in order to test existing models \cite{*[{See, e.g., }][] Strandburg88}. Despite the term of surface melting is originally introduced for equilibrium system, and most investigations are performed with thermodynamic equilibrium as a precondition, there exists evidence showing surface melting persists as a crystal is driven away from thermal equilibrium \cite{Dash99, Haekkinen93}. An interesting follow-up question is: Can our current microscopic view on melting be extended to the wide spreading non-equilibrium systems in nature?

Here, we try to address this question with a wet granular model system. Granular matter, besides its ubiquity in nature\cite{Duran00, Nagel96}, has been frequently used as a model system for phase transitions far from thermal equilibrium \cite{Pouligny90, Strassburger00, Olafsen05, Reis06, Fingerle08, Clerc08, Huang09a, Vazquez09, Roeller11, Castillo12}, due to the strongly dissipative particle-particle interactions. Here, we use a mono-layer of wet particles as a model system, because the cohesion arising from the formation of capillary bridges, which mimics molecular bonds, effectively leads to a crystalline structure with a free surface. The melting of such a wet granular crystal is found to be a two step process: A plastic deformation into an amorphous state, followed by melting from the surface. The abrupt transition into the surface melting regime is reminiscent of a first order phase transition. The transition threshold can be rationalized by a balance between the effective energy injection and the rupture energy of a single capillary bridge. 

%%%%%%%%%%%%%%%%%%%%%%%%%%%%%%%%%%%%%%%%%%%%%%%%%%%%%%%%%%
%%%%%%%%%%%%%%%%%%%%   Method   %%%%%%%%%%%%%%%%%%%%%%%%%%
%%%%%%%%%%%%%%%%%%%%%%%%%%%%%%%%%%%%%%%%%%%%%%%%%%%%%%%%%%

\begin{figure*}
  \centering
  \includegraphics[width = 0.85\textwidth]{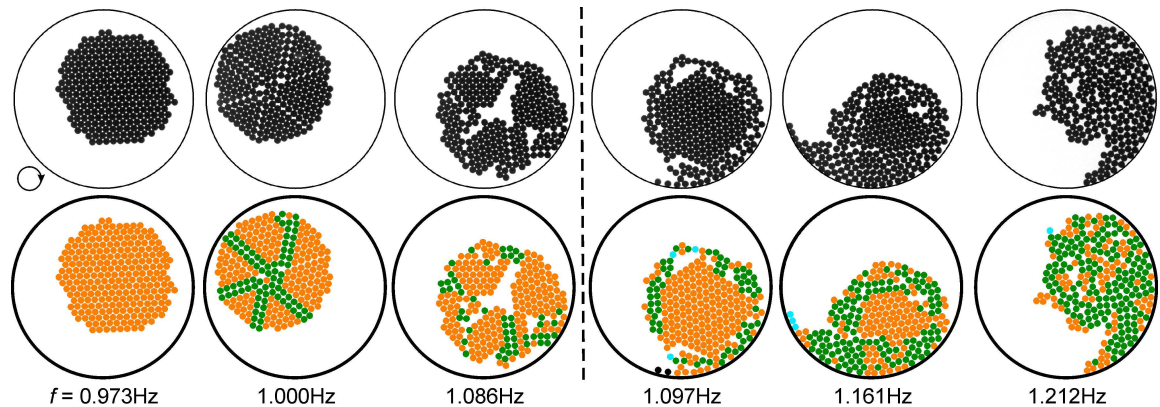}
  \caption{(color online) Melting process of a wet granular crystal with liquid content $W=2.4\%$ as the swirling frequency increases, represented by the snapshots (upper panel) and local structures (lower panel). The particles in the lower panel are color coded according to their local structures: Free particles with black; Line, square, hexagonal structures with blue, olive and red correspondingly. The gray dash line corresponds to the start of surface melting. The container swirls in the clockwise direction.}
  \label{fig:snapshot}
\end{figure*}
  
The wet granular sample is prepared by adding a certain volume of purified water $V_{\rm l}$ into a mono-layer of $N=250$ cleaned black glass spheres (SiliBeads P) with a density $\rho_{\rm p}=2.58~{\rm g/cm^{3}}$ and a diameter $d=4\pm0.02$~mm. The liquid content is defined as $W=V_{\rm l}/V_{\rm s}$ with $V_{\rm s}$ the total volume of the spheres. The cylindrical container made of polytetrafluoroethylene (PTFE) has an inner diameter of $D=102$~mm, a height of $6$~mm, and a glass lid sealed with Indium to avoid evaporation. The glass lid is heated slightly during the experiments to minimize the liquid condensation. The container is fixed on a swirling table leveled within $5.7\cdot10^{-3}$ degrees to avoid the influence from gravity. The swirling table (see \cite{Auma03} for a sketch) provides a horizontal circular motion with a frequency $f$ and amplitude $a$ as control parameters. This two combined horizontal vibrations provide an isotropic energy injection, since the amplitude of the agitation velocity is independent on the phase. The computer controlled swirling frequency can be varied with an accuracy of $6.2\cdot10^{-4}$~Hz. The dynamics of the spheres are captured by a camera (Lumenera LU125M) mounted on the co-moving frame of the swirling table. $f$ is obtained via tracing a fixed point on the swirling table with a second camera (Lumenera LU075M). The snapshots captured are subjected to an image processing procedure to locate all spheres based on a Hough transformation \cite{Kimme75}. 

From the positions found, the connectivity of two neighboring particles is determined through a comparison of their distance to the critical bond length $r_{\rm b}=1.25\,d$, which is estimated from the rupture distance of a capillary bridge \cite{Willett00}. Local symmetries of particle configurations are characterized with the bond orientational order parameters (BOOP) \cite{Steinhardt83, Wang05}. It is defined as 

\begin{equation}
\label{bop}
q_{\rm l}=\sqrt{\frac{4\pi}{2l+1}\sum_{m=-l}^{l}|\bar{Q}_{\rm lm}|^2},
\end{equation}

\noindent where $\bar{Q}_{\rm lm}\equiv \langle Q_{\rm lm}(\vec{r})\rangle$ is an average of the local order parameter $Q_{\rm lm}(\vec{r})\equiv Y_{\rm lm}(\theta(\vec{r}),\phi(\vec{r}))$ over all bonds connecting one particle to its nearest neighbors, with $Y_{\rm lm}(\theta(\vec{r}),\phi(\vec{r}))$ spherical harmonics of a bond located at $\vec{r}$. Here, we choose $q_{\rm 6}$ as the order parameter because of its sensitivity to the hexagonal order. Based on the deviations of $q_{\rm 6}$ to the standard values for perfectly hexagonal, square and line structures, the structure that a particle is most likely belonging to is distinguished.

%%%%%%%%%%%%%%%%%%%%%%%%%%%%%%%%%%%%%%%%%%%%%%%%%%%%%%%%%%
%%%%%%%%%%%%%%%%%%%% Result: initialization  %%%%%%%%%%%%
%%%%%%%%%%%%%%%%%%%%%%%%%%%%%%%%%%%%%%%%%%%%%%%%%%%%%%%%%%

To achieve a homogeneous wetting condition, the sample is swirled with a relatively large initial frequency for at least one hour. As agitation starts, the initially isolated particles merge with each other into small assemblies in a rather short time scale (few seconds), due to the cohesion arising from the formation of capillary bridges. As time evolves, those small assemblies gradually merge with each other into a single large cluster, within which the particles vigorously exchange positions with their neighbors, exhibiting a liquidlike state. To achieve an initial crystalline state, the swirling frequency is ramped down until the cluster stops reorganizing. Depending on the ramping rate, the initialized crystal may range from a perfectly hexagonal structure to a polycrystalline structure. In order to have a well defined initial condition, we keep the ramping rate slow enough for the system to favor the former structure.

%%%%%%%%%%%%%%%%%%%%%%%%%%%%%%%%%%%%%%%%%%%%%%%%%%%%%%%%%%
%%%%%%%%%%%%%%%%%%%% Result: melting  %%%%%%%%%%%%%%%%%%%%
%%%%%%%%%%%%%%%%%%%%%%%%%%%%%%%%%%%%%%%%%%%%%%%%%%%%%%%%%%

Figure~\ref{fig:snapshot} shows the melting process as the swirling frequency grows with a step of $6.2\cdot10^{-4}$~Hz and a waiting time of $1$ minute between each step. A variation of either parameter by one order of magnitude yield the same melting threshold. At the initial frequency $f=0.973$~Hz, the particles form a perfectly hexagonal structure. Although the crystal moves around collectively in the co-moving frame, the internal structure keeps stable. As the frequency increases to $1.000$~Hz, the occasional impacts with the container give rise to temporally formed cracks inside the crystal. Although the cracks formed fluctuate with time in such a non-equilibrium steady state, the reduced overall packing density and the weakened internal structure persist from a statistical point of view. Note that this is different from non-cohesive particles under vertical \cite{Huang06} or horizontal \cite{Strassburger00} agitations, where collisions with the container will tend to ``heat'' up the boundary layer and give rise to a granular temperature gradient as the ``heat'' flux propagates through. This difference could be attributed to the strong cohesion between adjacent particles, which leads to the favor collective motion. As the frequency increases further to $1.086$~Hz, more broken bonds lead to larger voids within the cluster, along with the plastic deformation of the crystal into an amorphous state. The enhanced energy dissipation at the defects effectively increases the susceptibility for the cluster to deform under normal stress applied by the container. Meanwhile, the shear stress drives the rotation of the cluster along the swirling direction.

An abrupt change of the structure occurs between $1.086$~Hz and $1.097$~Hz: All the voids inside the cluster disappear suddenly and a state with a perfectly hexagonal core surrounded by few liquidlike layers arises. We identify the new state as surface melting, since the deviations from the hexagonal structure locate only at the outer layers. At the beginning of the surface melting state, the liquidlike layer tends to ``wet'' the crystalline core and keeps a circular shape, suggesting a tendency to reduce its surface energy. As the frequency increases further to $1.161$~Hz, melting continues inward and the molten layer tends to deform along the swirling direction. Eventually at $1.212$~Hz, the sample melts completely into a comma shaped ``droplet'' moving along the rim of the container in the co-moving frame. The fluctuations of the surface of the molten layer, in connection to the interfacial tension of such a wet granular model system, is an interesting question to be addressed in further investigations.

\begin{figure}
  \centering
  \includegraphics[width = 0.45\textwidth]{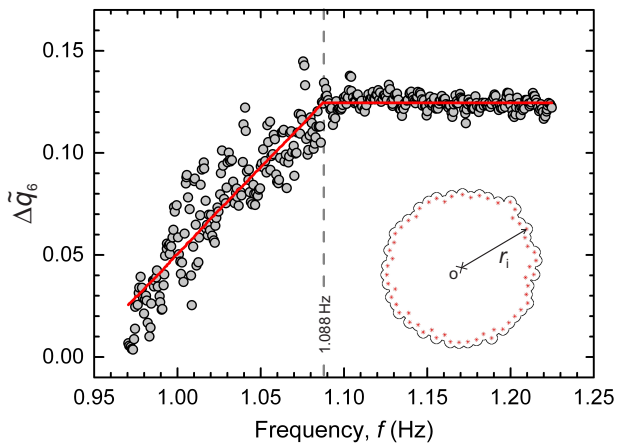}
  \caption{(color online) Deviation of the local structure for edge particles from a hexagonal one $\Delta \tilde{q}_{\rm 6}$ (see text for a definition) as frequency increases. Inset: a processed image with edge particles (stars) determined from the distance to the cluster contour line. $r_{\rm i}$ is the distance between an edge particle and the cluster center ${\rm O}$. The gray dash line corresponds to the melting threshold $f_{\rm th}=1.088\pm0.036$~Hz, which is determined by linear fits of the data. }
\label{fig:edge}
\end{figure}

To have a quantitative characterization of surface melting, we analyze the local structure of particles on the edge of a melting crystal, which is distinguished by the connectivity of particles: Based on the criteria described above, we locate all the neighbors of a particle, and recursively the neighbors of all the neighbors found, until all particles inside are found. The rescaled BOOP $\Delta \tilde{q_{\rm 6}}=|q_{\rm 6}/q_{\rm 6}^{\rm hex}-1|$ is chosen as the order parameter for surface melting, because it measures the deviation from a perfect hexagonal structure $q_{\rm 6}^{\rm hex}=0.741$. Note that $\Delta \tilde{q_{\rm 6}}=0$ corresponds to the initial crystalline state. Each data point corresponds to an average of all edge particles and over all frames captured.

As shown in Fig.~\ref{fig:edge}, surface melting can be clearly distinguished from the order parameter. Within data scattering, $\Delta \tilde{q}_{\rm 6}$ initially grows with the swirling frequency. This arises from the temporarily formed cracks across the crystal, as well as the following plastic deformation into the amorphous state, because both processes lead to a weakening of the cluster at the edge. As surface melting starts, $\Delta \tilde{q}_{\rm 6}$ saturates at a value of roughly $0.125$, since particles in a liquidlike state share a similar local structure. This behavior provides a convenient way to accurately determine $f_{\rm th}$. As the solid lines in Fig.~\ref{fig:edge} demonstrate, two fits, a 1st followed by a 0th order, are applied to the data, and the threshold corresponds to the intersection point that minimizes the standard error.

\begin{figure}
  \centering
  \includegraphics[width = 0.4\textwidth]{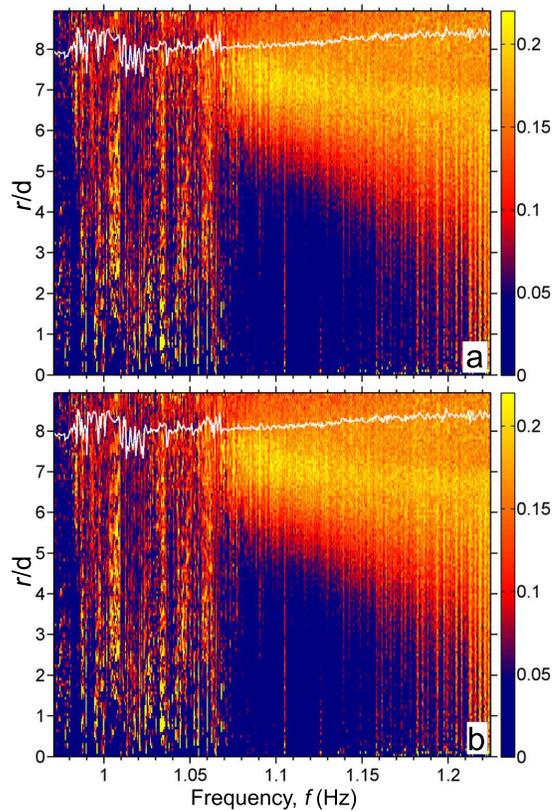}
  \caption{(color online) $\Delta \tilde{q}_{\rm 6}$ as a function of the rescaled distance $r/d$ to the cluster center ${\rm O}$ as the driving frequency, $f$, increases (a) and decreases(b). Here, $\Delta \tilde{q}_{\rm 6}$ corresponds to an average of the rescaled $q_{\rm 6}$ over all azimuth directions and all frames. The white curve in either plot corresponds to a measure of the cluster size $\bar{r}_{\rm i}$, which is averaged over all edge particles and all frames recorded.}
\label{fig:melting}
\end{figure}

Figure~\ref{fig:melting} (a) shows the internal structure of the crystal during the melting process, from which more features on the self-organization of particles inside the cluster can be obtained. First, it represents how the wet granular crystal evolves from crack forming to the amorphous state. In case of a crack with a constant width formed across the  crystal, it will lead to a deviation of $\Delta \tilde{q}_{\rm 6}$ for all particles associated. The deviation is pronounced in a periodic manner along the radial direction, because the positions of particles in such a crystal are mostly fixed to a hexagonal lattice. The decay of the deviation with $r$ arises from the average over all azimuth directions: The relative influence from the crack decreases as the distance to the center ${\rm O}$ increases. As $f$ grows, the enhanced crack formation leads to a larger deviation of $\Delta \tilde{q}_{\rm 6}$ and a continuous increase of the mean crystal size $\bar{r}_{\rm i}$. At about $1.016$~Hz, the deviation is strong enough to destroy the initial hexagonal lattice and allow a plastic deformation of the cluster. Consequently, large fluctuations of $\Delta \tilde{q}_{\rm 6}$ and of the cluster size start. Second, the change of internal structure presents the abrupt reorganization of the particles as surface melting starts. The amorphous state with large fluctuations of $\Delta \tilde{q}_{\rm 6}$ is suddenly replaced with a phase separation into a more compact inner core with a size of about $6d$, shielded with a molten layer of about $2d$, suggesting a first order like phase transition. Obvious deviations from a hexagonal packing can be observed within the molten layer, and its thickness grows monotonically with the driving frequency, along with the dilation of the cluster. Third, it indicates that the transition into the liquidlike state is not continuous: As the thickness of the molten layer reaches roughly half of the cluster size, the inner core of the cluster loses the hexagonal structure altogether and the whole cluster reaches a liquidlike state.

Figure~\ref{fig:melting} (b) shows the crystallization process as $f$ decreases with the same rate. Its similarity to the melting process is remarkable, except for a shift of the amorphous state to a slightly lower frequency. It suggests the existence of hysteresis between the melting and the crystallization process. Further experiments with various ramping rates up to $14$ runs indicate that the hysteresis, despite comparable to the error of the threshold, is reproducible. This behavior presumably arises from the hysteresis nature between the formation and rupture of a single capillary bridge \cite{Herminghaus05}.

%%%%%%%%%%%%%%%%%%%%%%%%%%%%%%%%%%%%%%%%%%%%%%%%%%%%%%%%%%
%%%%%%%%%%%%%%%%%%%% Model  %%%%%%%%%%%%%%%%%%%%%%%%%%%%%%
%%%%%%%%%%%%%%%%%%%%%%%%%%%%%%%%%%%%%%%%%%%%%%%%%%%%%%%%%%

\begin{figure}
  \centering
  \includegraphics[width = 0.45\textwidth]{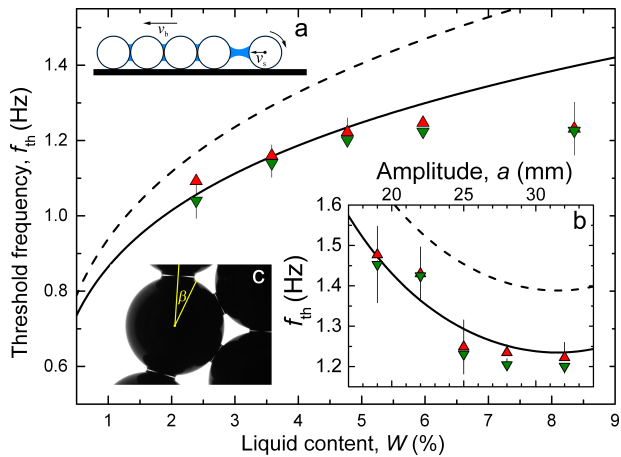}
  \caption{(color online) Dependency of the threshold frequency on the liquid content and the swirling amplitude (inset b). $f_{\rm th}$ is measured with the $\Delta \tilde{q}_{\rm 6}$ of edge particles as order parameter. The upper and lower triangles correspond to increasing and decreasing $f$. $a$ and $W$ are fixed at $31.8$~mm and at $W=4.77\%$ in the main panel and in b, correspondingly. The dash and solid lines are estimations from the model. Inset a: A sketch illustrating an edge particle rolling away from the bulk. $v_{\rm s}$ and $v_{\rm b}$ are the velocity of edge, and of the bulk particles in the lab frame. Inset c: A close view of the edge of a crystal with $W=2.39\%$, captured with a microscope. $\beta$ denotes the half opening angle of a capillary bridge.}
\label{fig:model}
\end{figure}

To gain further insights into the melting transition, the $f_{\rm th}$ is measured for various liquid content $W$ and swirling amplitude $a$. As shown in Fig.~\ref{fig:model}, the threshold increases monotonically with the liquid content and saturates at $W\approx6\%$. Assuming that surface melting represents the establishment of a new balance between the energy injection and dissipation, the dependence on $W$ can be rationalized with the enhanced energy dissipation due to the larger rupture distance of a capillary bridge. The viscous effect can be safely ignored, because the capillary number, the ratio between the viscous and capillary forces, is less than $10^{-2}$. The velocity of the swirling table $v_{\rm 0}=2\pi fa$, which plays a crucial role injecting energy into the system, provides a clue to understand the decay of $f_{\rm th}$ with the increase of $a$ shown in Fig.~\ref{fig:model}(b).

Following the above analysis, we propose a model based on a balance between the effective thermal energy injection $E_{\rm i}$ and the rupture energy of a single capillary bridge $E_{\rm b}$. The energy injection is considered to be a two step process: Colliding with the container wall provides `macroscopic' collective motion of the cluster, followed by a transfer into the `microscopic' thermal energy through particle-particle interactions inside. As illustrated in Fig.~\ref{fig:model}(a), the particle on the edge of a cluster has more freedom to roll compared with those in the bulk, due to less restrictions from the neighbors. This difference provides the thermal energy for the edge particle $m(2\pi fak)^2/2$, with $m$ its mass and $k=1-v_{\rm s}/v_{\rm b}$ the relative velocity difference. The latter is $5/7$ for the case that only edge particles roll, and roll without sliding and rolling frictions \cite{Kondic99}. As the second step only occurs without interruptions from the wall, we introduce an additional factor $\alpha=(D-2a)/(D+2a)$, the length scale for a particle to move without disturbance from the wall over that for the swirling table to reach. Therefore, the effective energy injection $E_{\rm i}=\alpha m(2\pi fak)^2/2$. On the other hand, the rupture energy of a capillary bridge can be estimated \cite{Willett00} to be $E_{\rm b}=3.68 \cos(\theta) \sigma \sqrt{V_{\rm b}d}$, with $\theta=0.227$ the contact angle obtained from a close view of the bridges, $\sigma=0.072$~N/m surface tension of water and the bridge volume $V_{\rm b}=\pi d^3 W/(3N_{\rm cor})$. The coordination number $N_{\rm cor}\approx 5.5$ is obtained from the initial crystalline state.

Consequently, the threshold frequency can be estimated with 

\begin{equation}
\label{fth}
f_{\rm th}=\frac{0.60}{ka}(\frac{\sigma \cos(\theta)}{\rho_{\rm p}d}\cdot\frac{D+2a}{D-2a})^{1/2}(\frac{W}{N_{\rm cor}})^{1/4},
\end{equation}

\noindent which is shown as dash lines in Fig.~\ref{fig:model}. The solid lines correspond to a more accurate estimation of $E_{\rm b}$ from a numerical integration of a more exact force law (appendix of \cite{Willett00}), which is accurate within $3\%$ for $W$ up to $6.58\%$. This limit corresponds to the merge of liquid bridges at $\beta=\pi/6$. A comparison with the experimental results indicate that, without any fit parameter, the model captures fairly well the dependency of $f_{\rm th}$ on $W$ and $A$, provided that the particles are connected via capillary bridges. The saturation of $f_{\rm th}$ with $W$ appears earlier than the limiting value, because the bridge volume is not always homogeneously distributed, as the snapshot shown in Fig.~\ref{fig:model} demonstrates. 

%%%%%%%%%%%%%%%%%%%%%%%%%%%%%%%%%%%%%%%%%%%%%%%%%%%%%%%%%%
%%%%%%%%%%%%%%%%%%%% Summary %%%%%%%%%%%%%%%%%%%%%%%%%%%%%
%%%%%%%%%%%%%%%%%%%%%%%%%%%%%%%%%%%%%%%%%%%%%%%%%%%%%%%%%%

In summary, the melting of a two dimensional wet granular crystal with 250 macroscopic ``molecules'' is demonstrated to be a continuous process starting from the surface. Preceding to surface melting, there exists an intermediate stage where the crystal deforms plastically into an amorphous state, leading to a more fragile internal structure with large fluctuations of voids. The abrupt transition into the surface melting regime, reminiscent of a first order phase transition, can be rationalized with the balance of the energy injection and dissipation through the rupture of capillary bridges. Moreover, this experiment indicates that gravity is not a crucial factor for the surface melting of such a model system, in connection to a former numerical investigation \cite{Roeller11} on veritically agitated wet granular matter. 

In the future, further investigations on the distribution of the granular temperature, especially close to the melting transition, are necessary to address the question on the possibility to extend our current microscopic view of surface melting into systems out of thermal equilibrium. A comparison to computer simulations will shed light on such an interesting question. Moreover, the tendency for the molten layer to minimize its free surface \cite{Lipowsky86, Tartaglino05, Cheng07, Luu13} provides the opportunity to investigate the interfacial tension of fluidized cohesive granular matter.
  
The authors would like to thank T.\ Fischer, A.\ Fortini, D.\ Nelson, J.\ Olafson, M.\ Schmidt and J.\ Vollmer for helpful discussions. This work is supported by DFG through grant No.~HU1939/2-1.

%\bibliography{2Dmelting}

\begin{thebibliography}{41}%
\makeatletter
\providecommand \@ifxundefined [1]{%
 \@ifx{#1\undefined}
}%
\providecommand \@ifnum [1]{%
 \ifnum #1\expandafter \@firstoftwo
 \else \expandafter \@secondoftwo
 \fi
}%
\providecommand \@ifx [1]{%
 \ifx #1\expandafter \@firstoftwo
 \else \expandafter \@secondoftwo
 \fi
}%
\providecommand \natexlab [1]{#1}%
\providecommand \enquote  [1]{``#1''}%
\providecommand \bibnamefont  [1]{#1}%
\providecommand \bibfnamefont [1]{#1}%
\providecommand \citenamefont [1]{#1}%
\providecommand \href@noop [0]{\@secondoftwo}%
\providecommand \href [0]{\begingroup \@sanitize@url \@href}%
\providecommand \@href[1]{\@@startlink{#1}\@@href}%
\providecommand \@@href[1]{\endgroup#1\@@endlink}%
\providecommand \@sanitize@url [0]{\catcode `\\12\catcode `\$12\catcode
  `\&12\catcode `\#12\catcode `\^12\catcode `\_12\catcode `\%12\relax}%
\providecommand \@@startlink[1]{}%
\providecommand \@@endlink[0]{}%
\providecommand \url  [0]{\begingroup\@sanitize@url \@url }%
\providecommand \@url [1]{\endgroup\@href {#1}{\urlprefix }}%
\providecommand \urlprefix  [0]{URL }%
\providecommand \Eprint [0]{\href }%
\providecommand \doibase [0]{http://dx.doi.org/}%
\providecommand \selectlanguage [0]{\@gobble}%
\providecommand \bibinfo  [0]{\@secondoftwo}%
\providecommand \bibfield  [0]{\@secondoftwo}%
\providecommand \translation [1]{[#1]}%
\providecommand \BibitemOpen [0]{}%
\providecommand \bibitemStop [0]{}%
\providecommand \bibitemNoStop [0]{.\EOS\space}%
\providecommand \EOS [0]{\spacefactor3000\relax}%
\providecommand \BibitemShut  [1]{\csname bibitem#1\endcsname}%
\let\auto@bib@innerbib\@empty
%</preamble>
\bibitem [{\citenamefont {Faraday}(1850)}]{Faraday1850}%
  \BibitemOpen
  \bibfield  {author} {\bibinfo {author} {\bibfnamefont {M.}~\bibnamefont
  {Faraday}},\ }\href@noop {} {\bibfield  {journal} {\bibinfo  {journal} {The
  Athenaeum}\ }\textbf {\bibinfo {volume} {1181}},\ \bibinfo {pages} {640}
  (\bibinfo {year} {1850})}\BibitemShut {NoStop}%
\bibitem [{\citenamefont {Dash}(1999)}]{Dash99}%
  \BibitemOpen
  \bibfield  {author} {\bibinfo {author} {\bibfnamefont {J.~G.}\ \bibnamefont
  {Dash}},\ }\href {\doibase 10.1103/RevModPhys.71.1737} {\bibfield  {journal}
  {\bibinfo  {journal} {Rev. Mod. Phys.}\ }\textbf {\bibinfo {volume} {71}},\
  \bibinfo {pages} {1737} (\bibinfo {year} {1999})}\BibitemShut {NoStop}%
\bibitem [{\citenamefont {Tartaglino}\ \emph {et~al.}(2005)\citenamefont
  {Tartaglino}, \citenamefont {Zykova-Timan}, \citenamefont {Ercolessi},\ and\
  \citenamefont {Tosatti}}]{Tartaglino05}%
  \BibitemOpen
  \bibfield  {author} {\bibinfo {author} {\bibfnamefont {U.}~\bibnamefont
  {Tartaglino}}, \bibinfo {author} {\bibfnamefont {T.}~\bibnamefont
  {Zykova-Timan}}, \bibinfo {author} {\bibfnamefont {F.}~\bibnamefont
  {Ercolessi}}, \ and\ \bibinfo {author} {\bibfnamefont {E.}~\bibnamefont
  {Tosatti}},\ }\href {\doibase DOI: 10.1016/j.physrep.2005.01.004} {\bibfield
  {journal} {\bibinfo  {journal} {Physics Reports}\ }\textbf {\bibinfo {volume}
  {411}},\ \bibinfo {pages} {291 } (\bibinfo {year} {2005})}\BibitemShut
  {NoStop}%
\bibitem [{\citenamefont {Dash}\ \emph {et~al.}(2006)\citenamefont {Dash},
  \citenamefont {Rempel},\ and\ \citenamefont {Wettlaufer}}]{Dash06}%
  \BibitemOpen
  \bibfield  {author} {\bibinfo {author} {\bibfnamefont {J.~G.}\ \bibnamefont
  {Dash}}, \bibinfo {author} {\bibfnamefont {A.~W.}\ \bibnamefont {Rempel}}, \
  and\ \bibinfo {author} {\bibfnamefont {J.~S.}\ \bibnamefont {Wettlaufer}},\
  }\href {\doibase 10.1103/RevModPhys.78.695} {\bibfield  {journal} {\bibinfo
  {journal} {Rev. Mod. Phys.}\ }\textbf {\bibinfo {volume} {78}},\ \bibinfo
  {pages} {695} (\bibinfo {year} {2006})}\BibitemShut {NoStop}%
\bibitem [{\citenamefont {Frenkel}(1946)}]{Frenkel46}%
  \BibitemOpen
  \bibfield  {author} {\bibinfo {author} {\bibfnamefont {J.}~\bibnamefont
  {Frenkel}},\ }\href@noop {} {\emph {\bibinfo {title} {Kinetic Theory of
  Liquids}}}\ (\bibinfo  {publisher} {Clarendon, Oxford},\ \bibinfo {year}
  {1946})\BibitemShut {NoStop}%
\bibitem [{\citenamefont {Frenken}\ and\ \citenamefont
  {Veen}(1985)}]{Frenken85}%
  \BibitemOpen
  \bibfield  {author} {\bibinfo {author} {\bibfnamefont {J.~W.~M.}\
  \bibnamefont {Frenken}}\ and\ \bibinfo {author} {\bibfnamefont {J.~F. v.~d.}\
  \bibnamefont {Veen}},\ }\href {\doibase 10.1103/PhysRevLett.54.134}
  {\bibfield  {journal} {\bibinfo  {journal} {Phys. Rev. Lett.}\ }\textbf
  {\bibinfo {volume} {54}},\ \bibinfo {pages} {134} (\bibinfo {year}
  {1985})}\BibitemShut {NoStop}%
\bibitem [{\citenamefont {Frenken}\ \emph {et~al.}(1986)\citenamefont
  {Frenken}, \citenamefont {Mar\'ee},\ and\ \citenamefont {van~der
  Veen}}]{Frenken86}%
  \BibitemOpen
  \bibfield  {author} {\bibinfo {author} {\bibfnamefont {J.~W.~M.}\
  \bibnamefont {Frenken}}, \bibinfo {author} {\bibfnamefont {P.~M.~J.}\
  \bibnamefont {Mar\'ee}}, \ and\ \bibinfo {author} {\bibfnamefont {J.~F.}\
  \bibnamefont {van~der Veen}},\ }\href {\doibase 10.1103/PhysRevB.34.7506}
  {\bibfield  {journal} {\bibinfo  {journal} {Phys. Rev. B}\ }\textbf {\bibinfo
  {volume} {34}},\ \bibinfo {pages} {7506} (\bibinfo {year}
  {1986})}\BibitemShut {NoStop}%
\bibitem [{\citenamefont {Cahn}(1986)}]{Cahn86}%
  \BibitemOpen
  \bibfield  {author} {\bibinfo {author} {\bibfnamefont {R.~W.}\ \bibnamefont
  {Cahn}},\ }\href {\doibase 10.1038/323668a0} {\bibfield  {journal} {\bibinfo
  {journal} {Nature}\ }\textbf {\bibinfo {volume} {323}},\ \bibinfo {pages}
  {668} (\bibinfo {year} {1986})}\BibitemShut {NoStop}%
\bibitem [{\citenamefont {Jin}\ \emph {et~al.}(2001)\citenamefont {Jin},
  \citenamefont {Gumbsch}, \citenamefont {Lu},\ and\ \citenamefont
  {Ma}}]{Jin01}%
  \BibitemOpen
  \bibfield  {author} {\bibinfo {author} {\bibfnamefont {Z.~H.}\ \bibnamefont
  {Jin}}, \bibinfo {author} {\bibfnamefont {P.}~\bibnamefont {Gumbsch}},
  \bibinfo {author} {\bibfnamefont {K.}~\bibnamefont {Lu}}, \ and\ \bibinfo
  {author} {\bibfnamefont {E.}~\bibnamefont {Ma}},\ }\href {\doibase
  10.1103/PhysRevLett.87.055703} {\bibfield  {journal} {\bibinfo  {journal}
  {Phys. Rev. Lett.}\ }\textbf {\bibinfo {volume} {87}},\ \bibinfo {pages}
  {055703} (\bibinfo {year} {2001})}\BibitemShut {NoStop}%
\bibitem [{\citenamefont {Cahn}(2001)}]{Cahn01}%
  \BibitemOpen
  \bibfield  {author} {\bibinfo {author} {\bibfnamefont {R.~W.}\ \bibnamefont
  {Cahn}},\ }\href {\doibase 10.1038/35098169} {\bibfield  {journal} {\bibinfo
  {journal} {Nature}\ }\textbf {\bibinfo {volume} {413}},\ \bibinfo {pages}
  {582} (\bibinfo {year} {2001})}\BibitemShut {NoStop}%
\bibitem [{\citenamefont {Zahn}\ and\ \citenamefont {Maret}(2000)}]{Zahn00}%
  \BibitemOpen
  \bibfield  {author} {\bibinfo {author} {\bibfnamefont {K.}~\bibnamefont
  {Zahn}}\ and\ \bibinfo {author} {\bibfnamefont {G.}~\bibnamefont {Maret}},\
  }\href {\doibase 10.1103/PhysRevLett.85.3656} {\bibfield  {journal} {\bibinfo
   {journal} {Phys. Rev. Lett.}\ }\textbf {\bibinfo {volume} {85}},\ \bibinfo
  {pages} {3656} (\bibinfo {year} {2000})}\BibitemShut {NoStop}%
\bibitem [{\citenamefont {Gasser}\ \emph {et~al.}(2001)\citenamefont {Gasser},
  \citenamefont {Weeks}, \citenamefont {Schofield}, \citenamefont {Pusey},\
  and\ \citenamefont {Weitz}}]{Gasser2001}%
  \BibitemOpen
  \bibfield  {author} {\bibinfo {author} {\bibfnamefont {U.}~\bibnamefont
  {Gasser}}, \bibinfo {author} {\bibfnamefont {E.~R.}\ \bibnamefont {Weeks}},
  \bibinfo {author} {\bibfnamefont {A.}~\bibnamefont {Schofield}}, \bibinfo
  {author} {\bibfnamefont {P.~N.}\ \bibnamefont {Pusey}}, \ and\ \bibinfo
  {author} {\bibfnamefont {D.~A.}\ \bibnamefont {Weitz}},\ }\href {\doibase
  10.1126/science.1058457} {\bibfield  {journal} {\bibinfo  {journal}
  {Science}\ }\textbf {\bibinfo {volume} {292}},\ \bibinfo {pages} {258}
  (\bibinfo {year} {2001})}\BibitemShut {NoStop}%
\bibitem [{\citenamefont {Alsayed}\ \emph {et~al.}(2005)\citenamefont
  {Alsayed}, \citenamefont {Islam}, \citenamefont {Zhang}, \citenamefont
  {Collings},\ and\ \citenamefont {Yodh}}]{Alsayed2005}%
  \BibitemOpen
  \bibfield  {author} {\bibinfo {author} {\bibfnamefont {A.~M.}\ \bibnamefont
  {Alsayed}}, \bibinfo {author} {\bibfnamefont {M.~F.}\ \bibnamefont {Islam}},
  \bibinfo {author} {\bibfnamefont {J.}~\bibnamefont {Zhang}}, \bibinfo
  {author} {\bibfnamefont {P.~J.}\ \bibnamefont {Collings}}, \ and\ \bibinfo
  {author} {\bibfnamefont {A.~G.}\ \bibnamefont {Yodh}},\ }\href {\doibase
  10.1126/science.1112399} {\bibfield  {journal} {\bibinfo  {journal}
  {Science}\ }\textbf {\bibinfo {volume} {309}},\ \bibinfo {pages} {1207}
  (\bibinfo {year} {2005})}\BibitemShut {NoStop}%
\bibitem [{\citenamefont {Gasser}\ \emph {et~al.}(2010)\citenamefont {Gasser},
  \citenamefont {Eisenmann}, \citenamefont {Maret},\ and\ \citenamefont
  {Keim}}]{Gasser10}%
  \BibitemOpen
  \bibfield  {author} {\bibinfo {author} {\bibfnamefont {U.}~\bibnamefont
  {Gasser}}, \bibinfo {author} {\bibfnamefont {C.}~\bibnamefont {Eisenmann}},
  \bibinfo {author} {\bibfnamefont {G.}~\bibnamefont {Maret}}, \ and\ \bibinfo
  {author} {\bibfnamefont {P.}~\bibnamefont {Keim}},\ }\href {\doibase
  10.1002/cphc.201090021} {\bibfield  {journal} {\bibinfo  {journal}
  {ChemPhysChem}\ }\textbf {\bibinfo {volume} {11}},\ \bibinfo {pages} {925}
  (\bibinfo {year} {2010})}\BibitemShut {NoStop}%
\bibitem [{\citenamefont {Wang}\ \emph {et~al.}(2012)\citenamefont {Wang},
  \citenamefont {Wang}, \citenamefont {Peng}, \citenamefont {Zheng},\ and\
  \citenamefont {Han}}]{Wang12}%
  \BibitemOpen
  \bibfield  {author} {\bibinfo {author} {\bibfnamefont {Z.}~\bibnamefont
  {Wang}}, \bibinfo {author} {\bibfnamefont {F.}~\bibnamefont {Wang}}, \bibinfo
  {author} {\bibfnamefont {Y.}~\bibnamefont {Peng}}, \bibinfo {author}
  {\bibfnamefont {Z.}~\bibnamefont {Zheng}}, \ and\ \bibinfo {author}
  {\bibfnamefont {Y.}~\bibnamefont {Han}},\ }\href {\doibase
  10.1126/science.1224763} {\bibfield  {journal} {\bibinfo  {journal}
  {Science}\ }\textbf {\bibinfo {volume} {338}},\ \bibinfo {pages} {87}
  (\bibinfo {year} {2012})}\BibitemShut {NoStop}%
\bibitem [{\citenamefont {Strandburg}(1988)}]{Strandburg88}%
  \BibitemOpen
  \bibfield  {author} {\bibinfo {author} {\bibfnamefont {K.~J.}\ \bibnamefont
  {Strandburg}},\ }\href {\doibase 10.1103/RevModPhys.60.161} {\bibfield
  {journal} {\bibinfo  {journal} {Rev. Mod. Phys.}\ }\textbf {\bibinfo {volume}
  {60}},\ \bibinfo {pages} {161} (\bibinfo {year} {1988})}\BibitemShut
  {NoStop}%
\bibitem [{\citenamefont {Hakkinen}\ and\ \citenamefont
  {Landman}(1993)}]{Haekkinen93}%
  \BibitemOpen
  \bibfield  {author} {\bibinfo {author} {\bibfnamefont {H.}~\bibnamefont
  {Hakkinen}}\ and\ \bibinfo {author} {\bibfnamefont {U.}~\bibnamefont
  {Landman}},\ }\href {\doibase 10.1103/PhysRevLett.71.1023} {\bibfield
  {journal} {\bibinfo  {journal} {Phys. Rev. Lett.}\ }\textbf {\bibinfo
  {volume} {71}},\ \bibinfo {pages} {1023} (\bibinfo {year}
  {1993})}\BibitemShut {NoStop}%
\bibitem [{\citenamefont {Duran}(2000)}]{Duran00}%
  \BibitemOpen
  \bibfield  {author} {\bibinfo {author} {\bibfnamefont {J.}~\bibnamefont
  {Duran}},\ }\href@noop {} {\emph {\bibinfo {title} {Sands, Powders and Grains
  (An Introduction to the Physics of Granular Materials)}}},\ \bibinfo
  {edition} {1st}\ ed.\ (\bibinfo  {publisher} {Springer-Verlag},\ \bibinfo
  {address} {New York, Inc.},\ \bibinfo {year} {2000})\BibitemShut {NoStop}%
\bibitem [{\citenamefont {Jaeger}\ \emph {et~al.}(1996)\citenamefont {Jaeger},
  \citenamefont {Nagel},\ and\ \citenamefont {Behringer}}]{Nagel96}%
  \BibitemOpen
  \bibfield  {author} {\bibinfo {author} {\bibfnamefont {H.~M.}\ \bibnamefont
  {Jaeger}}, \bibinfo {author} {\bibfnamefont {S.~R.}\ \bibnamefont {Nagel}}, \
  and\ \bibinfo {author} {\bibfnamefont {R.~P.}\ \bibnamefont {Behringer}},\
  }\href {\doibase 10.1103/RevModPhys.68.1259} {\bibfield  {journal} {\bibinfo
  {journal} {Rev. Mod. Phys.}\ }\textbf {\bibinfo {volume} {68}},\ \bibinfo
  {pages} {1259} (\bibinfo {year} {1996})}\BibitemShut {NoStop}%
\bibitem [{\citenamefont {Pouligny}\ \emph {et~al.}(1990)\citenamefont
  {Pouligny}, \citenamefont {Malzbender}, \citenamefont {Ryan},\ and\
  \citenamefont {Clark}}]{Pouligny90}%
  \BibitemOpen
  \bibfield  {author} {\bibinfo {author} {\bibfnamefont {B.}~\bibnamefont
  {Pouligny}}, \bibinfo {author} {\bibfnamefont {R.}~\bibnamefont
  {Malzbender}}, \bibinfo {author} {\bibfnamefont {P.}~\bibnamefont {Ryan}}, \
  and\ \bibinfo {author} {\bibfnamefont {N.~A.}\ \bibnamefont {Clark}},\ }\href
  {\doibase 10.1103/PhysRevB.42.988} {\bibfield  {journal} {\bibinfo  {journal}
  {Phys. Rev. B}\ }\textbf {\bibinfo {volume} {42}},\ \bibinfo {pages} {988}
  (\bibinfo {year} {1990})}\BibitemShut {NoStop}%
\bibitem [{\citenamefont {Strassburger}\ and\ \citenamefont
  {Rehberg}(2000)}]{Strassburger00}%
  \BibitemOpen
  \bibfield  {author} {\bibinfo {author} {\bibfnamefont {G.}~\bibnamefont
  {Strassburger}}\ and\ \bibinfo {author} {\bibfnamefont {I.}~\bibnamefont
  {Rehberg}},\ }\href {\doibase 10.1103/PhysRevE.62.2517} {\bibfield  {journal}
  {\bibinfo  {journal} {Phys. Rev. E}\ }\textbf {\bibinfo {volume} {62}},\
  \bibinfo {pages} {2517} (\bibinfo {year} {2000})}\BibitemShut {NoStop}%
\bibitem [{\citenamefont {Olafsen}\ and\ \citenamefont
  {Urbach}(2005)}]{Olafsen05}%
  \BibitemOpen
  \bibfield  {author} {\bibinfo {author} {\bibfnamefont {J.~S.}\ \bibnamefont
  {Olafsen}}\ and\ \bibinfo {author} {\bibfnamefont {J.~S.}\ \bibnamefont
  {Urbach}},\ }\href {\doibase 10.1103/PhysRevLett.95.098002} {\bibfield
  {journal} {\bibinfo  {journal} {Phys. Rev. Lett.}\ }\textbf {\bibinfo
  {volume} {95}},\ \bibinfo {pages} {098002} (\bibinfo {year}
  {2005})}\BibitemShut {NoStop}%
\bibitem [{\citenamefont {Reis}\ \emph {et~al.}(2006)\citenamefont {Reis},
  \citenamefont {Ingale},\ and\ \citenamefont {Shattuck}}]{Reis06}%
  \BibitemOpen
  \bibfield  {author} {\bibinfo {author} {\bibfnamefont {P.~M.}\ \bibnamefont
  {Reis}}, \bibinfo {author} {\bibfnamefont {R.~A.}\ \bibnamefont {Ingale}}, \
  and\ \bibinfo {author} {\bibfnamefont {M.~D.}\ \bibnamefont {Shattuck}},\
  }\href {\doibase 10.1103/PhysRevLett.96.258001} {\bibfield  {journal}
  {\bibinfo  {journal} {Phys. Rev. Lett.}\ }\textbf {\bibinfo {volume} {96}},\
  \bibinfo {pages} {258001} (\bibinfo {year} {2006})}\BibitemShut {NoStop}%
\bibitem [{\citenamefont {Fingerle}\ \emph {et~al.}(2008)\citenamefont
  {Fingerle}, \citenamefont {Roeller}, \citenamefont {Huang},\ and\
  \citenamefont {Herminghaus}}]{Fingerle08}%
  \BibitemOpen
  \bibfield  {author} {\bibinfo {author} {\bibfnamefont {A.}~\bibnamefont
  {Fingerle}}, \bibinfo {author} {\bibfnamefont {K.}~\bibnamefont {Roeller}},
  \bibinfo {author} {\bibfnamefont {K.}~\bibnamefont {Huang}}, \ and\ \bibinfo
  {author} {\bibfnamefont {S.}~\bibnamefont {Herminghaus}},\ }\href@noop {}
  {\bibfield  {journal} {\bibinfo  {journal} {New J. Phys.}\ }\textbf {\bibinfo
  {volume} {10}},\ \bibinfo {pages} {053020} (\bibinfo {year}
  {2008})}\BibitemShut {NoStop}%
\bibitem [{\citenamefont {Clerc}\ \emph {et~al.}(2008)\citenamefont {Clerc},
  \citenamefont {Cordero}, \citenamefont {Dunstan}, \citenamefont {Huff},
  \citenamefont {Mujica}, \citenamefont {Risso},\ and\ \citenamefont
  {Varas}}]{Clerc08}%
  \BibitemOpen
  \bibfield  {author} {\bibinfo {author} {\bibfnamefont {M.~G.}\ \bibnamefont
  {Clerc}}, \bibinfo {author} {\bibfnamefont {P.}~\bibnamefont {Cordero}},
  \bibinfo {author} {\bibfnamefont {J.}~\bibnamefont {Dunstan}}, \bibinfo
  {author} {\bibfnamefont {K.}~\bibnamefont {Huff}}, \bibinfo {author}
  {\bibfnamefont {N.}~\bibnamefont {Mujica}}, \bibinfo {author} {\bibfnamefont
  {D.}~\bibnamefont {Risso}}, \ and\ \bibinfo {author} {\bibfnamefont
  {G.}~\bibnamefont {Varas}},\ }\href {\doibase 10.1038/nphys884} {\bibfield
  {journal} {\bibinfo  {journal} {Nat Phys}\ }\textbf {\bibinfo {volume} {4}},\
  \bibinfo {pages} {249} (\bibinfo {year} {2008})}\BibitemShut {NoStop}%
\bibitem [{\citenamefont {Huang}\ \emph {et~al.}(2009)\citenamefont {Huang},
  \citenamefont {Roeller},\ and\ \citenamefont {Herminghaus}}]{Huang09a}%
  \BibitemOpen
  \bibfield  {author} {\bibinfo {author} {\bibfnamefont {K.}~\bibnamefont
  {Huang}}, \bibinfo {author} {\bibfnamefont {K.}~\bibnamefont {Roeller}}, \
  and\ \bibinfo {author} {\bibfnamefont {S.}~\bibnamefont {Herminghaus}},\
  }\href@noop {} {\bibfield  {journal} {\bibinfo  {journal} {Eur. Phys. J.
  Special Topics}\ }\textbf {\bibinfo {volume} {179}},\ \bibinfo {pages} {25}
  (\bibinfo {year} {2009})}\BibitemShut {NoStop}%
\bibitem [{\citenamefont {Pacheco-V\'azquez}\ \emph {et~al.}(2009)\citenamefont
  {Pacheco-V\'azquez}, \citenamefont {Caballero-Robledo},\ and\ \citenamefont
  {Ruiz-Su\'arez}}]{Vazquez09}%
  \BibitemOpen
  \bibfield  {author} {\bibinfo {author} {\bibfnamefont {F.}~\bibnamefont
  {Pacheco-V\'azquez}}, \bibinfo {author} {\bibfnamefont {G.~A.}\ \bibnamefont
  {Caballero-Robledo}}, \ and\ \bibinfo {author} {\bibfnamefont {J.~C.}\
  \bibnamefont {Ruiz-Su\'arez}},\ }\href {\doibase
  10.1103/PhysRevLett.102.170601} {\bibfield  {journal} {\bibinfo  {journal}
  {Phys. Rev. Lett.}\ }\textbf {\bibinfo {volume} {102}},\ \bibinfo {pages}
  {170601} (\bibinfo {year} {2009})}\BibitemShut {NoStop}%
\bibitem [{\citenamefont {Roeller}\ and\ \citenamefont
  {Herminghaus}(2011)}]{Roeller11}%
  \BibitemOpen
  \bibfield  {author} {\bibinfo {author} {\bibfnamefont {K.}~\bibnamefont
  {Roeller}}\ and\ \bibinfo {author} {\bibfnamefont {S.}~\bibnamefont
  {Herminghaus}},\ }\href {\doibase 10.1209/0295-5075/96/26003} {\bibfield
  {journal} {\bibinfo  {journal} {Europhys. Lett.}\ }\textbf {\bibinfo {volume}
  {96}},\ \bibinfo {pages} {26003} (\bibinfo {year} {2011})}\BibitemShut
  {NoStop}%
\bibitem [{\citenamefont {Castillo}\ \emph {et~al.}(2012)\citenamefont
  {Castillo}, \citenamefont {Mujica},\ and\ \citenamefont {Soto}}]{Castillo12}%
  \BibitemOpen
  \bibfield  {author} {\bibinfo {author} {\bibfnamefont {G.}~\bibnamefont
  {Castillo}}, \bibinfo {author} {\bibfnamefont {N.}~\bibnamefont {Mujica}}, \
  and\ \bibinfo {author} {\bibfnamefont {R.}~\bibnamefont {Soto}},\ }\href
  {\doibase 10.1103/PhysRevLett.109.095701} {\bibfield  {journal} {\bibinfo
  {journal} {Phys. Rev. Lett.}\ }\textbf {\bibinfo {volume} {109}},\ \bibinfo
  {pages} {095701} (\bibinfo {year} {2012})}\BibitemShut {NoStop}%
\bibitem [{\citenamefont {Auma\^\i{}tre}\ \emph {et~al.}(2003)\citenamefont
  {Auma\^\i{}tre}, \citenamefont {Schnautz}, \citenamefont {Kruelle},\ and\
  \citenamefont {Rehberg}}]{Auma03}%
  \BibitemOpen
  \bibfield  {author} {\bibinfo {author} {\bibfnamefont {S.}~\bibnamefont
  {Auma\^\i{}tre}}, \bibinfo {author} {\bibfnamefont {T.}~\bibnamefont
  {Schnautz}}, \bibinfo {author} {\bibfnamefont {C.~A.}\ \bibnamefont
  {Kruelle}}, \ and\ \bibinfo {author} {\bibfnamefont {I.}~\bibnamefont
  {Rehberg}},\ }\href {\doibase 10.1103/PhysRevLett.90.114302} {\bibfield
  {journal} {\bibinfo  {journal} {Phys. Rev. Lett.}\ }\textbf {\bibinfo
  {volume} {90}},\ \bibinfo {pages} {114302} (\bibinfo {year}
  {2003})}\BibitemShut {NoStop}%
\bibitem [{\citenamefont {Kimme}\ \emph {et~al.}(1975)\citenamefont {Kimme},
  \citenamefont {Ballard},\ and\ \citenamefont {Sklansky}}]{Kimme75}%
  \BibitemOpen
  \bibfield  {author} {\bibinfo {author} {\bibfnamefont {C.}~\bibnamefont
  {Kimme}}, \bibinfo {author} {\bibfnamefont {D.}~\bibnamefont {Ballard}}, \
  and\ \bibinfo {author} {\bibfnamefont {J.}~\bibnamefont {Sklansky}},\
  }\href@noop {} {\bibfield  {journal} {\bibinfo  {journal} {Communications of
  the {ACM}}\ }\textbf {\bibinfo {volume} {18}},\ \bibinfo {pages} {120}
  (\bibinfo {year} {1975})}\BibitemShut {NoStop}%
\bibitem [{\citenamefont {Willett}\ \emph {et~al.}(2000)\citenamefont
  {Willett}, \citenamefont {Adams}, \citenamefont {Johnson},\ and\
  \citenamefont {Seville}}]{Willett00}%
  \BibitemOpen
  \bibfield  {author} {\bibinfo {author} {\bibfnamefont {C.~D.}\ \bibnamefont
  {Willett}}, \bibinfo {author} {\bibfnamefont {M.~J.}\ \bibnamefont {Adams}},
  \bibinfo {author} {\bibfnamefont {S.~A.}\ \bibnamefont {Johnson}}, \ and\
  \bibinfo {author} {\bibfnamefont {J.~P.~K.}\ \bibnamefont {Seville}},\ }\href
  {\doibase 10.1021/la000657y} {\bibfield  {journal} {\bibinfo  {journal}
  {Langmuir}\ }\textbf {\bibinfo {volume} {16}},\ \bibinfo {pages} {9396}
  (\bibinfo {year} {2000})}\BibitemShut {NoStop}%
\bibitem [{\citenamefont {Steinhardt}\ \emph {et~al.}(1983)\citenamefont
  {Steinhardt}, \citenamefont {Nelson},\ and\ \citenamefont
  {Ronchetti}}]{Steinhardt83}%
  \BibitemOpen
  \bibfield  {author} {\bibinfo {author} {\bibfnamefont {P.~J.}\ \bibnamefont
  {Steinhardt}}, \bibinfo {author} {\bibfnamefont {D.~R.}\ \bibnamefont
  {Nelson}}, \ and\ \bibinfo {author} {\bibfnamefont {M.}~\bibnamefont
  {Ronchetti}},\ }\href@noop {} {\bibfield  {journal} {\bibinfo  {journal}
  {Phys. Rev. B}\ }\textbf {\bibinfo {volume} {28}},\ \bibinfo {pages} {784}
  (\bibinfo {year} {1983})}\BibitemShut {NoStop}%
\bibitem [{\citenamefont {Wang}\ \emph {et~al.}(2005)\citenamefont {Wang},
  \citenamefont {Teitel},\ and\ \citenamefont {Dellago}}]{Wang05}%
  \BibitemOpen
  \bibfield  {author} {\bibinfo {author} {\bibfnamefont {Y.}~\bibnamefont
  {Wang}}, \bibinfo {author} {\bibfnamefont {S.}~\bibnamefont {Teitel}}, \ and\
  \bibinfo {author} {\bibfnamefont {C.}~\bibnamefont {Dellago}},\ }\href@noop
  {} {\bibfield  {journal} {\bibinfo  {journal} {J. Chem. Phys.}\ }\textbf
  {\bibinfo {volume} {122}},\ \bibinfo {pages} {214722} (\bibinfo {year}
  {2005})}\BibitemShut {NoStop}%
\bibitem [{\citenamefont {Huang}\ \emph {et~al.}(2006)\citenamefont {Huang},
  \citenamefont {Miao}, \citenamefont {Zhang}, \citenamefont {Yun},\ and\
  \citenamefont {Wei}}]{Huang06}%
  \BibitemOpen
  \bibfield  {author} {\bibinfo {author} {\bibfnamefont {K.}~\bibnamefont
  {Huang}}, \bibinfo {author} {\bibfnamefont {G.}~\bibnamefont {Miao}},
  \bibinfo {author} {\bibfnamefont {P.}~\bibnamefont {Zhang}}, \bibinfo
  {author} {\bibfnamefont {Y.}~\bibnamefont {Yun}}, \ and\ \bibinfo {author}
  {\bibfnamefont {R.}~\bibnamefont {Wei}},\ }\href {\doibase
  10.1103/PhysRevE.73.041302} {\bibfield  {journal} {\bibinfo  {journal} {Phys.
  Rev. E}\ }\textbf {\bibinfo {volume} {73}},\ \bibinfo {pages} {041302}
  (\bibinfo {year} {2006})}\BibitemShut {NoStop}%
\bibitem [{\citenamefont {Herminghaus}(2005)}]{Herminghaus05}%
  \BibitemOpen
  \bibfield  {author} {\bibinfo {author} {\bibfnamefont {S.}~\bibnamefont
  {Herminghaus}},\ }\href {\doibase 10.1080/00018730500167855} {\bibfield
  {journal} {\bibinfo  {journal} {Adv. Phys.}\ }\textbf {\bibinfo {volume}
  {54}},\ \bibinfo {pages} {221} (\bibinfo {year} {2005})}\BibitemShut
  {NoStop}%
\bibitem [{\citenamefont {Kondic}(1999)}]{Kondic99}%
  \BibitemOpen
  \bibfield  {author} {\bibinfo {author} {\bibfnamefont {L.}~\bibnamefont
  {Kondic}},\ }\href {\doibase 10.1103/PhysRevE.60.751} {\bibfield  {journal}
  {\bibinfo  {journal} {Phys. Rev. E}\ }\textbf {\bibinfo {volume} {60}},\
  \bibinfo {pages} {751} (\bibinfo {year} {1999})}\BibitemShut {NoStop}%
\bibitem [{\citenamefont {Lipowsky}(1986)}]{Lipowsky86}%
  \BibitemOpen
  \bibfield  {author} {\bibinfo {author} {\bibfnamefont {R.}~\bibnamefont
  {Lipowsky}},\ }\href {\doibase 10.1103/PhysRevLett.57.2876} {\bibfield
  {journal} {\bibinfo  {journal} {Phys. Rev. Lett.}\ }\textbf {\bibinfo
  {volume} {57}},\ \bibinfo {pages} {2876} (\bibinfo {year}
  {1986})}\BibitemShut {NoStop}%
\bibitem [{\citenamefont {Cheng}\ \emph {et~al.}(2007)\citenamefont {Cheng},
  \citenamefont {Varas}, \citenamefont {Citron}, \citenamefont {Jaeger},\ and\
  \citenamefont {Nagel}}]{Cheng07}%
  \BibitemOpen
  \bibfield  {author} {\bibinfo {author} {\bibfnamefont {X.}~\bibnamefont
  {Cheng}}, \bibinfo {author} {\bibfnamefont {G.}~\bibnamefont {Varas}},
  \bibinfo {author} {\bibfnamefont {D.}~\bibnamefont {Citron}}, \bibinfo
  {author} {\bibfnamefont {H.~M.}\ \bibnamefont {Jaeger}}, \ and\ \bibinfo
  {author} {\bibfnamefont {S.~R.}\ \bibnamefont {Nagel}},\ }\href {\doibase
  10.1103/PhysRevLett.99.188001} {\bibfield  {journal} {\bibinfo  {journal}
  {Phys. Rev. Lett.}\ }\textbf {\bibinfo {volume} {99}},\ \bibinfo {pages}
  {188001} (\bibinfo {year} {2007})}\BibitemShut {NoStop}%
\bibitem [{\citenamefont {Luu}\ \emph {et~al.}(2013)\citenamefont {Luu},
  \citenamefont {Castillo}, \citenamefont {Mujica},\ and\ \citenamefont
  {Soto}}]{Luu13}%
  \BibitemOpen
  \bibfield  {author} {\bibinfo {author} {\bibfnamefont {L.-H.}\ \bibnamefont
  {Luu}}, \bibinfo {author} {\bibfnamefont {G.}~\bibnamefont {Castillo}},
  \bibinfo {author} {\bibfnamefont {N.}~\bibnamefont {Mujica}}, \ and\ \bibinfo
  {author} {\bibfnamefont {R.}~\bibnamefont {Soto}},\ }\href {\doibase
  10.1103/PhysRevE.87.040202} {\bibfield  {journal} {\bibinfo  {journal} {Phys.
  Rev. E}\ }\textbf {\bibinfo {volume} {87}},\ \bibinfo {pages} {040202}
  (\bibinfo {year} {2013})}\BibitemShut {NoStop}%
\end{thebibliography}

\end{document}